# EPIDERMAL CORNEOCYTES: DEAD GUARDS OF THE HIDDEN TREASURE

**AV Mezentsev**
*Center for Radiological Research, Columbia University NY, New York, USA*

Acknowledgements:

Abbreviations:
7-DHC, 7-dehydrocholesterol; VDR, vitamin D receptor; PLC, phospholipase C, $PIP_2$, phosphatidylinositols; $IP_3$, inositol triphosphate; DAG, diacylglycerol; TGM, transglutaminase(s); ALP, antileukoproteinase; LEKTI, lymphoepithelial Kazal-type 5 serine protease inhibitor; SCCE, corneum chymotryptic enzyme; SCTE, stratum corneum tryptic enzyme; MT-SP1, Matriptase.

## Abstract

Gradual transformation of the epidermal stem cells to corneocytes involves a chain of chronologically well-arranged events that mostly stimulated locally by their neighbors. Cell diversity that observed during the differentiation through the different epidermal cell layers included the consisted changes of cell shape, intercellular contacts and proliferation. However, the most dramatically these changes appeared at the molecular level through gene expression, catalysis and intraprotein interactions. The proposed review explains these changes by switching systemic transcription factors that unlike their counterparts those role is limited to a contribution to gene expression also prepare cells to the next step of differentiation *via* modification of the chromatin pattern . Since primary epidermal keratinocytes are one of the most easy available type of the stem cells, a better understanding of the epidermal differentiation will benefit the research in the other areas by a discovery of basic coordinating mechanisms that stand behind such distinct molecular events as cell signaling and gene expression, and formulate basic principles for a smart therapeutic correction of the metabolism.

## Introduction

As the most outer tissue of the body, the epidermis protects it from physical and chemical insults and infections. In the other words, the epidermis keeps inside what do we need and outside - whatever can harm us. It also maintains homeostasis through thermoregulation and prevents dehydration. Besides, the epidermis contributes to the immune surveillance and also employed as a sensory organ. Successful functioning of the epidermis is based on its ability to the continuous self-renewal that obviously requires a balance between cellular proliferation and cell death. Disturbance of the balance leads to life-threatening conditions: excessive cell proliferation causes a development of tumors and prevalence of the cell death causes the tissue atrophy. The barrier function of the epidermis is based on a balance between established cell junctions and protease inhibitors from one side and proteases- from another one. The necessity to connect cells in the internal layers requires different kind of cell-cell contacts. These areas are protected from degradation by protease inhibitors unless the arrest of gene expression that normally occurs in the upper layers will raise proteolytic activity and disrupts the cell junctions.

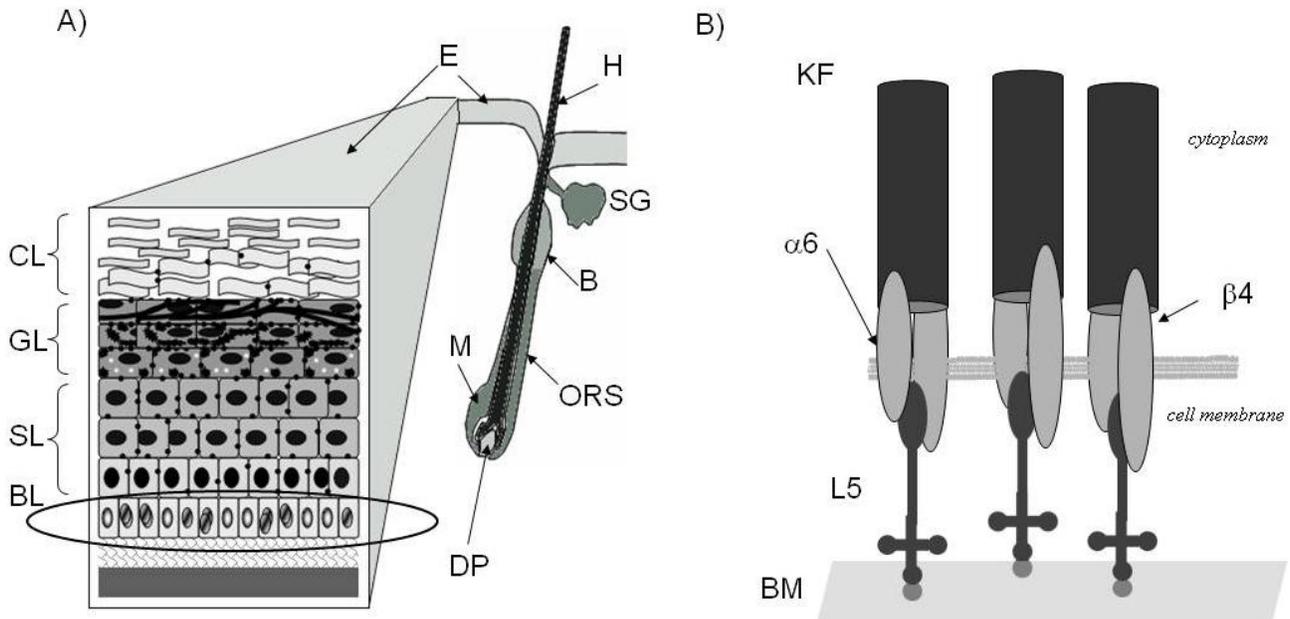

Figure 1 **A. Interfolicular epidermis and hair follicle: location of stem cells in the epidermis**. Stem cells are located within the bulge region of the hair follicle and at the interfollicular space of the basement membrane. In the picture: E-epidermis; H- hair; SG- sebaceous gland; B-bulge; M-matrix; ORS- outer root sheath; DP- dermal papilla; CL- cornified layer, *stratum corneum*; GL- granular layer, stratum granulosum; SL- spinous layer, *stratum spinosum*; BL- basal layer, *stratum basale*. **B. Hemi-desmosomes: connection of cells to the basal membrane**. In the picture: KF- keratin filaments; α6 and β4- α6 and β4 integrins; L5- laminin 5; BM- basal membrane.

## Proliferating cells of the basal membrane

Proliferating cells of a self-renewing tissue should maintain its thickness and prevent life-threatening changes in the genome. In the epidermis, this attained due to a coordinated work of two different kinds of proliferating cells: epidermal stem cells that retain genome integrity and transient amplifying cells that are responsible for the most of cell divisions. Stem cells are located within the bulge region of hair follicles and at interfollicular space of the basement membrane (Figure 1A). They are pluripotent cells that divide infrequently [1]. Proliferation of these slow-cycling cells occurs asymmetrically and gives a life to one new stem cell and one differentiating (transit amplifying) cell (Figure 2A). Unlike the stem cells, transit amplifying cells divide symmetrically and one parent cell produces two differentiating cells (Figure 2B). Until epidermal cells express integrins α6, β1 and β4 [2] they continue to proliferate. Mice conditionally null for β1 integrin exhibit an impaired proliferation in the epidermis and expose defects under the assembly of the basement membrane [3-5].

In *stratum basale*, cells are of cuboidal shape and maintain a spatio-temporal orientation. They are attached to the basal membrane through hemi-desmosomes. Hemi-desmosomes (Figure 1B) are made of laminin 5 that is anchored in the basal membrane, integrin α6β4 and intercellular keratin filaments [6] The α6β4 integrin-null mice show epidermal blistering, which is consistent with the role for α6β4 integrin in the anchorage [7]. Besides, cells located on the basal membrane receive signals from the extracellular matrix to regulate organization of the cytoskeleton, proliferation and differentiation [8].

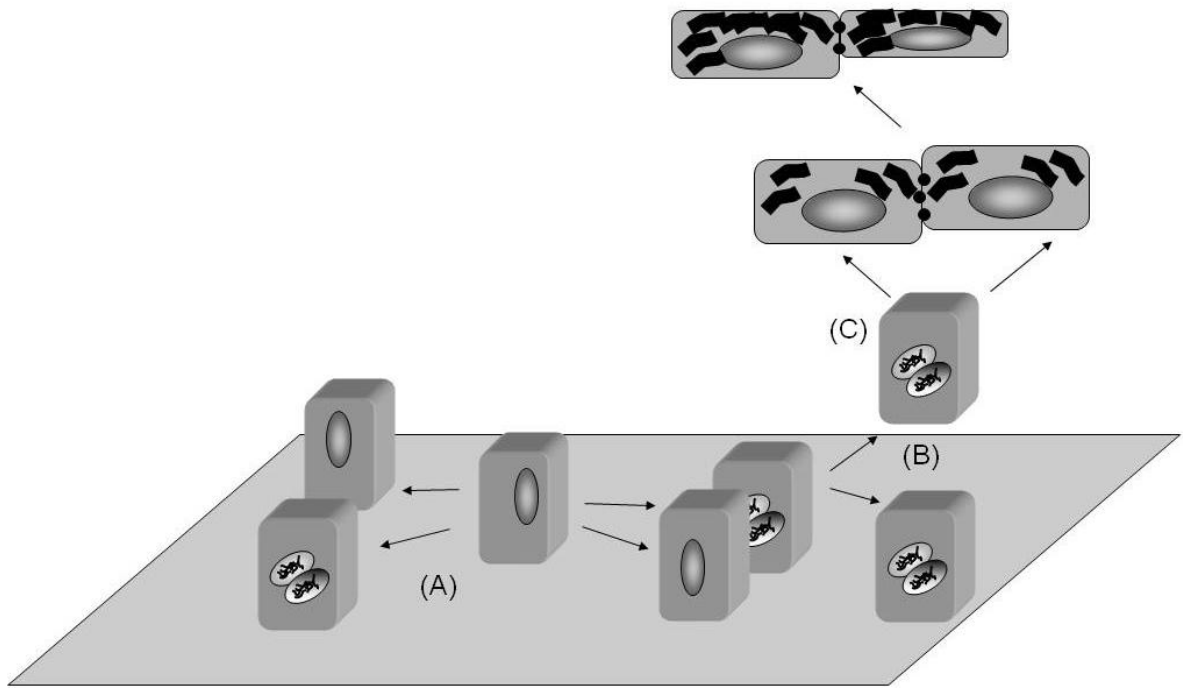

Figure 2 **Division and differentiation of epidermal stem and transit amplifying cells**. A. Dividing epidermal stem cell divides asymmetrically to one stem cell and one transit amplifying cell. B. Transit amplifying cell divides symmetrically to two differentiating cells; C. Differentiating cells loose a connection with the basal membrane, interrupt cell division and move to the spinous layer.

Analysis of the signaling pathways employed in differentiation will reveal what and how drives the stem cells to corneocytes. Unfortunately, the mechanisms that control population of stem cells in the epidermis are not yet uncovered [1]. However, two signaling pathways (Wnt and Myc) seem to be implicated in this process. The key element of the canonical Wnt signaling pathway, β-catenin, plays a dual role in the epidermis. First, β-catenin was identified as a part of adherent junctions between the cells of the basal membrane [9] where it links actin filaments of the cytoplasm and cadherins of the plasma membrane. The cytoplasmic pool of β-catenin is phosphorylated by the kinase, GSK-3, and degrades in the cell proteasomes (Figure 3A). Activation of Wnt pathway inhibits the phosphorylation and leads to a translocation of β-catenin to the nucleus where it interacts with transcription factor Lef/Tcf (Fig 3B). This complex binds to the DNA and activates the expression of numerous genes including sonic hedgehog (SHH). After secretion from the cells, SHH binds to the receptor Patched at the surface of another cell. Patched is an integral membrane protein that acts as an inhibitor of Smoothened activation. The pathway downstream of the Smoothened receptor remains unclear. However, it involves the Gli family of transcriptional activators, including Gli-1, Gli-2, and Gli-3 and induces the proliferative response (Figure 3C) [10]. Interestingly, interaction of β-catenin with the repressive form of Lef/Tcf overlaps with epigenetic control of gene expression since it displaces HDAC complex (Figure 3D) [11; 12]. Constitutive activation of β-catenin in the epidermis of transgenic mice leads to a highly enriched stem cell population. On the other hand, dominant-negative β-catenin favors cell differentiation and stimulates cells to leave the stem cell compartments [10].

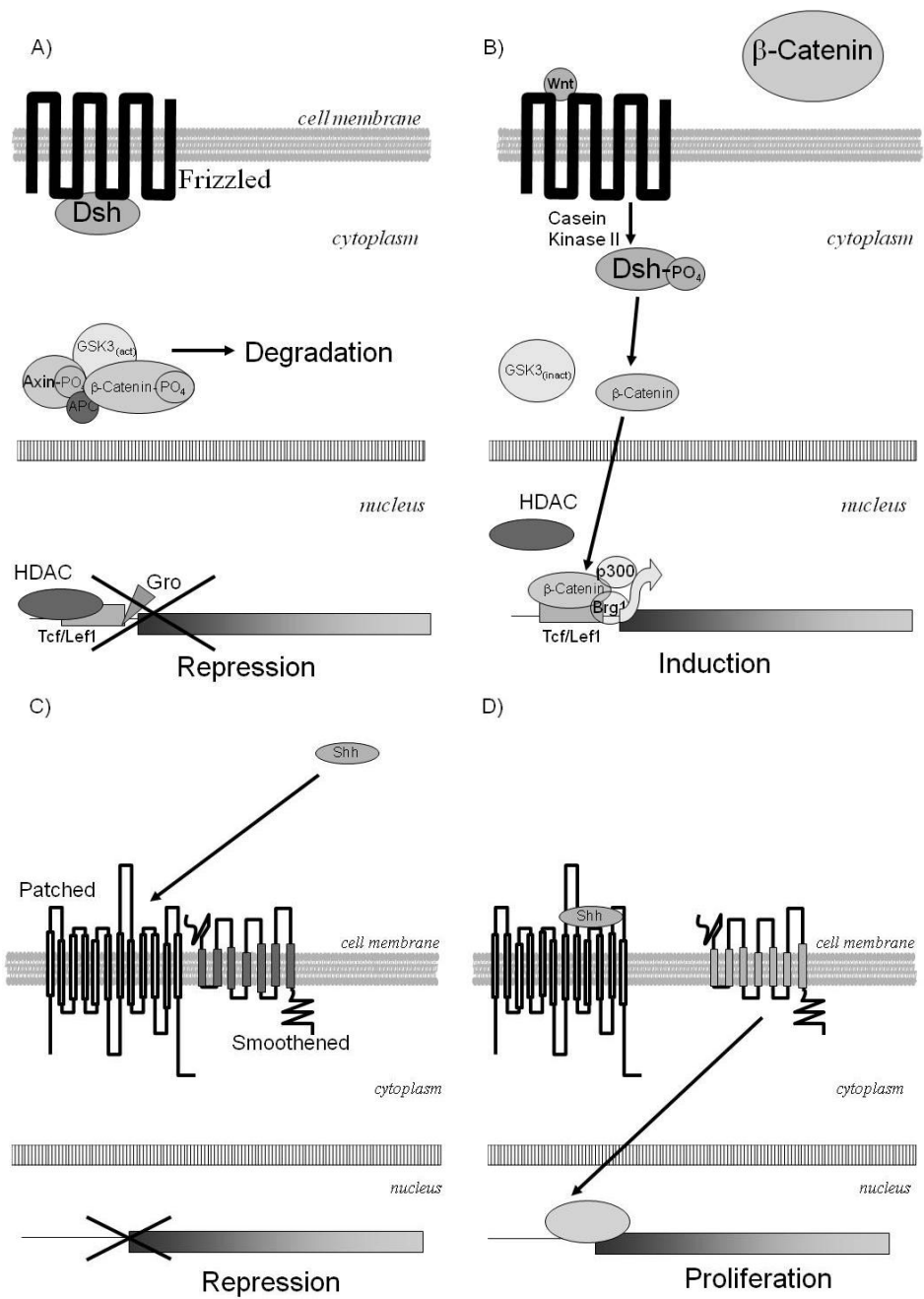

Figure 3    **Role of the canonical Wnt signaling pathway in gene expression and proliferation**.
A. Cell with deactivated Wnt signaling pathway: excess of free (unbound) β-catenin degrades in proteasomes. B. Cell after activation of Wnt-signaling pathway: β-catenin binds to Tcf/Lef factor, causing dissociation of the transcriptional repressors and induces the gene expression.

In the picture: β-cat, β-catenin; Dsh- mediator of Wnt signaling; Axin, Gsk3, and APC- subunits of the β-catenin degradation complex; Brg1- SWI/SNF chromatin-remodeling protein, repressor; p300-coactivator; HDAC, histone deacetylase, histone modifying enzyme, repressor; Gro, member of groucho family of protein repressors. C. Sonic hedgehog pathway: cell prior activation by SHH
D. Activation of sonic hedgehog pathway by SHH.

One of the β-catenin downstream targets is oncogene, c-Myc [13]. c-Myc is required for the transition of the cell from G1 to S phase of the cell cycle and it promotes proliferation of the epidermal transit amplifying cells. It was shown that activation of c-Myc pathway in the epidermis stimulates cell differentiation and departure from the basal membrane [14; 15]. The majority of the genes suppressed by c-Myc in the epidermis encodes cell adhesion and cytoskeleton proteins. Besides, c-Myc downregulates the expression of the α6β4 integrin, decreases the formation of hemi-desmosomes and delays the assembly of the actomyosin cytoskeleton. Level of c-Myc also has a reverse correlation with decreased levels of the other target proteins. Size and shape of the keratinocytes become variable and nuclear atypia appears as well [14; 15]. The skin looses melanocytes and Langerhans cells. Response to the injury [16], chemical clearance and immune response are also slower than in control [13;14]. The epidermis becomes more fragile and more frequently develops a cancer [13]. On the other hand, action of c-Myc, as it was shown recently, depends on the presence of another protein, Miz1. The transgenic mice with a lack of functional Myc-Miz1 heterodimer restored normal keratinocyte adhesion, polarization of the basal membrane despite of overexpression of c-Myc [17].

Conversion of stem cells to transit amplifying cells is likely to consider as the first step of keratinocyte differentiation. They still reside on the basal membrane, the innermost layer of the epidermis, and migrate on it laterally. However, their protein expression profile has already changed due to an induction of markers of early differentiation: keratins 5 and 14 (K5 and K14) and p63 [2].

## Keratinocytes in transition to the squamous cells

After they lose a connection to the basal membrane, the differentiating cells enter the suprabasal layer and become squamous cells (Figure 2C). Despite they experience similar external and intrinsic signals as their multipotent embryonic counterparts, the differentiating cells undergo changes in morphology and gene expression as well as gradual changes in their junctions and water impermeability. In *stratum spinosum*, cells increase their size and cytoplasm: nucleus ratio. They also gain an extensive cytoskeleton which is made of keratin filaments and linked to tightly adhesive *desmosomes* [18; 19]. Unlike the basal membrane that composed of the only layer of cells, the spinous layer is about four to six cells thick.

Cell proliferation in basal and suprabasal layers of the epidermis can be activated through vitamin D pathway by the sun light that converts 7-dehydrocholesterol (7-DHC) to vitamin D-3 in a photochemical reaction [20, 21]. The following hydroxylation of D-3 produces calcitriol, $1,25(OH)_2D_3$, the active form of vitamin D, that induces gene expression through the binding to vitamin D receptor (VDR). After the binding, the receptor forms either homo- or heterodimer with different binding partners: $VDR_2$; VDR-RXR or VDR-RAR in favor of VDR-RXR. The dimers translocate to the nucleus where they interact to the DNA binding either transcriptional repressor DRIP or coactivator p160/SRC at the DNA. The dimers translocate to the nucleus where they interact to the DNA binding either transcriptional repressor DRIP or coactivator p160/SRC. In undifferentiated keratinocytes, the dimers presumably bind to DRIP while in differentiating cells binding to p160/SRC takes over due to the level of DRIP falls down [*rev. in* 21]. Interestingly, higher doses of Vitamin D have a negative effect on the proliferation [compare 22 and 23]. Administration of $1,25(OH)_2D_3$ blocks the cells at the transition from $G_o/G_1$ to S phase of the cell cycle [24] due to an induction of the cell cycle inhibitor, p21 [25]. Influence of VDR signaling on keratinocyte differentiation was recently confirmed by characterization of VDR-null phenotype in the mice. VDR-null mice exhibit a defect in epidermal differentiation, it believed, due to reduced levels of involucrin, profilaggrin, and loricrin, and lack of keratohyalin granules in the epidermis [26].

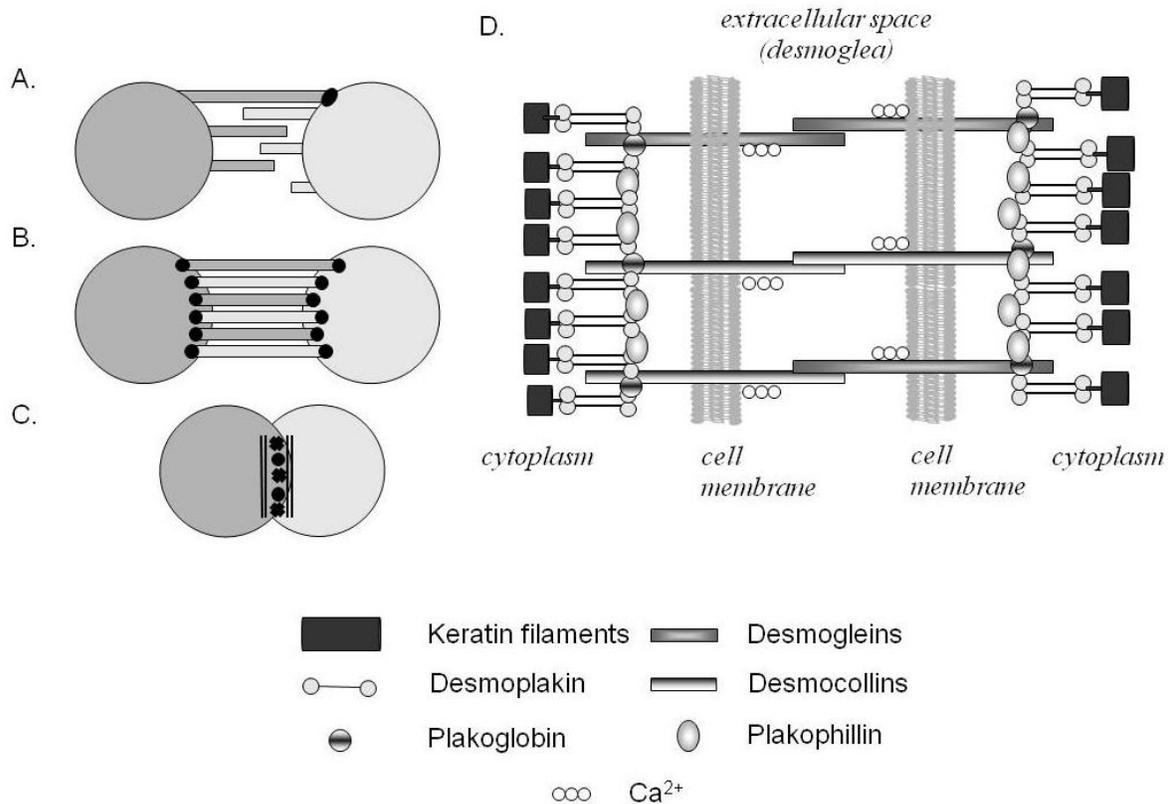

Figure 4    **Desmosomes in keratinocytes: assembly and schematic representation**. A. Merging keratinocytes form filopodia that embed into the neighbor cell. B. Formation of two-row adhesion zipper between the merging cells. C. Stabilization of the desmosomes. D. Parts of the mature desmosome and their interaction.

Activation of NFkB pathway drives the cells to the quiescent state [27]. Normally, it occurs in the suprabasal layer while in the cells of the basal membrane, NFκB remains inactive. Inhibition of NFκB leads to hyperproliferation of keratinocytes and thickening of the epidermis while its overexpression- thins the tissue and causes hypoproliferation [28].

One of the interesting phenomena observed in the spinous layer was the formation of desmosomes (Figure 4). Desmosomes are the molecular complexes of linking and cell adhesion proteins that connect cells to each other and secure the connection through the binding to the intermediate cytoskeletal filaments. In the epidermis, the desmosomes are composed of transmembrane desmosomal cadherins (desmocollin and desmoglein) and two linker proteins called desmoplakin and plakophilin. Desmoplakin attaches cytoplasmic parts of the cadherins to the intermediate filament network. There are two options that can be used by the cell to form a link between desmosomal cadherins and intrafilaments. Both involve C-terminal domain of the linker protein, desmoplakin [29; 30]. Desmoplakin either binds the filaments directly to desmocollin [31] or connects them to desmogleins through the binding to plakoglobin and plakophilin.

Formation of adherens junctions and desmosomes requires extracellular calcium. Raising the calcium concentration in the cell culture medium from 0.05 to 1.2mM [32] stimulates keratinocytes to form strong cell-cell adhesions *in vitro*. The adherent junctions appear in the cell prior the desmosomes [33] and are necessary for arrangement the last [34-37]. The calcium switch leads to a rapid reorganization of actin and intermediate filaments and relocation of junctional and desmosomal proteins towards cell-cell borders [33; 38]. Assembly of a desmosome starts from a generation of filopodia (Figure 4A). The filopodia penetrate and embed into the adjacent cell. Adherent junctional proteins concentrate at the tip of the filopodia and generate a two-row adhesion zipper (Figure 4B) [36]. Then, the desmosome binds the opposing cells together and stabilizes the junction [36; 37] until the directed actin polymerization will transform the two-row adhesion zipper into a single row (Figure 4C).

Moving upward through the spinous layer, the cells initiate the program of terminal differentiation. In *stratum spinosum*, keratinocytes express the following differentiation markers: K1 and K10, involucrin (Ivl), envoplakin, periplakin and 14-3-3σ, a nuclear export protein implicated in the Ras/MAPK pathway [39]. Expression of Ivl, envoplakin and periplakin appears to be the beginning of cornification.

## Keratinocytes in the granular layer, $Ca^{2+}$ influx and its role in differentiation

In the next epidermal level, *stratum granulosum*, cells have an elongated shape and contain multiple keratohyaline granules in the cytoplasm. These granules mostly consist of loricrin and profilaggrin, the precursor of filaggrin, a bundling protein of the keratin filaments. The cells of *stratum granulosum* also contain lamellar bodies mostly composed of lipids and lipid-processing enzymes whose contents will be secreted later into the junction between *stratum granulosum* and *stratum corneum* and become a mortar between the corneocyte "bricks". Usually, the granular layer composed of two- four layers of cells. Cells contain keratohyalin granules and number of the granules as well as their size increases toward the skin surface. The granules tend to accumulate in the outer apical side of the cytoplasm in each cell.

The majority of regulatory proteins employed in cornification depend on $Ca^{2+}$. Outside the cell, calcium concentration rises toward the stratum corneum and keratinocyte differentiation is tightly linked to the calcium extracellular contents [*rev. in* 21]. Upwelling extracellular calcium causes a calcium influx into the cell [40-42]. This influx is mediated by the calcium receptor (CaR, Figure 5) [43, 44], which is normally expressed in the suprabasal keratinocytes [45] and upregulated by the active form of vitamin D [46]. With progress of differentiation, keratinocytes lose their sensitivity to calcium [47], and this loss coincides with a switch in CaR processing: from the long to the alternatively spliced short one that lacks of one of the exons [43]. While several ion channels of the cell plasma membrane were identified as potential gates for the calcium invasion [48-52], their role has still to be characterized. Absence of the full-length CaR in the epidermis resulted in hyperproliferation and ultrastructural changes of the granular keratinocytes such as abnormal keratohyalin granule formation and premature secretion of lamellar body. Late-differentiation proteins (Flg, Lor and Ivl) were downregulated while number of proliferating cells was dramatically increased even in the presence of well-formed $Ca^{2+}$ gradient across the epidermis [45].

In the cells, calcium activates phospholipase C (PLC), particularly the isoforms PLCβ and PLCγ1 [53, 54]. PLC is the key enzyme of phosphoinositol phosphates metabolism (Figure 5) that hydrolyzes the membrane lipids, phosphatidylinositols ($PIP_2$), into two different secondary messengers: inositol triphosphate ($IP_3$) and diacylglycerol (DAG). The first compound, water-soluble $IP_3$ diffuses through the cytoplasm to the endoplasmic reticulum where it binds to and opens calcium channels, releasing calcium from the intercellular storages into the cytoplasm. Interaction with both: DAG and calcium activates PKC increasing the kinase activity. This leads to phosphorylation of many other proteins, including the transcription factor AP1, altering their activity. The AP1-dependent transcriptional activation upregulates the expression of the cornified envelope precursors and transglutaminases (TGM). This accelerates the formation of the cornified envelope.

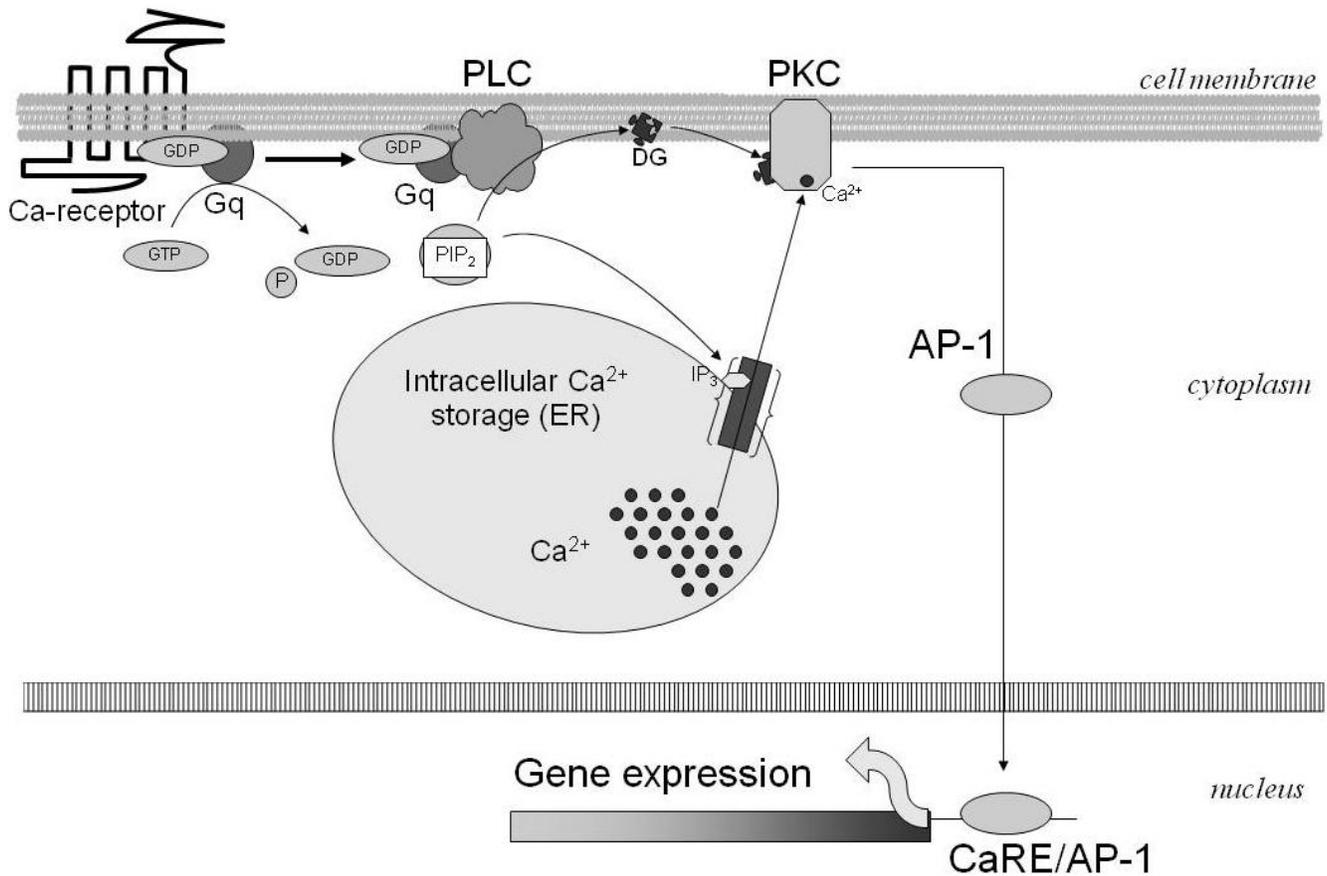

Figure 5. **Ca$^{2+}$-mediated signaling in keratinocytes.** Calcitriol, 1,25(OH)$_2$D$_3$, activates CaR and accelerates conversion of phosphatidylinositol bisphosphate to IP3, and DG. Their rise leads to activation of protein kinase (PKC) by translocating PKCs to their specific membrane receptors (RACK). Opening of calcium channels in the plasma membrane and the following downstream activation of AP-1 transcription factors induces expression of cornified envelope precursors and transglutaminase (TG) and leads to formation of cornified envelope (CE).

## Assembling of the cornified envelope

At the step of terminal differentiation, keratinocytes assemble the cornified cell envelope, an insoluble, about 10-nm thick protein layer (Figure 6E). In the epidermis, the cornified envelop also includes 5-nm thick lipid layer made of the ceramides that are covalently bound to the proteins [55; 56]. There are at least two types of chemical bonds that connect the components of the cornified envelope: N$^\varepsilon$- (γ-glutamyl)lysine isopeptide bonds produced during the crosslinking by transglutaminases and disulfide (S-S) bonds [55; 57]. Once the cornified envelop seals the body surface, the epidermal cells start carrying their barrier function [58].

Cornification starts in the upper part of the spinous layer with an induction of the early differentiation genes. At the stage, called "initiation", keratinocytes express envoplakin and periplakin (Figure 6A and B) [59]. This followed by the expression of involucrin that binds to the cell membranes in a Ca$^{2+}$-dependent manner [60] and members of S100 family: S100a7, S100a8, S100a9 and S100a10 [51]. While S100a8 and S100a9 can be considered as stress-induced proteins, S100a10 is interesting by its ability to interact with annexin II and form Ca$^{2+}$- channels in the plasma membrane promoting the cornification (Figure 6C).

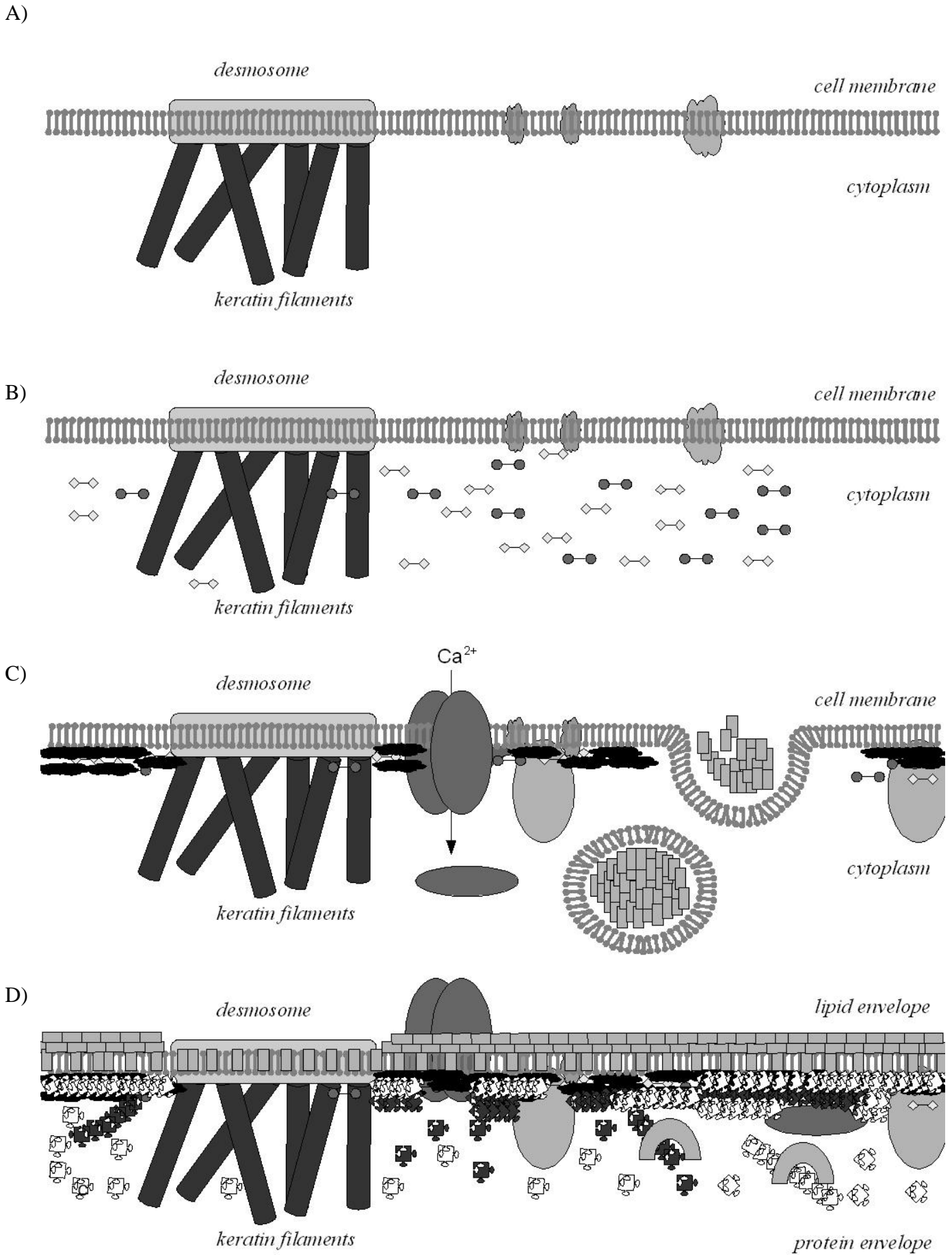

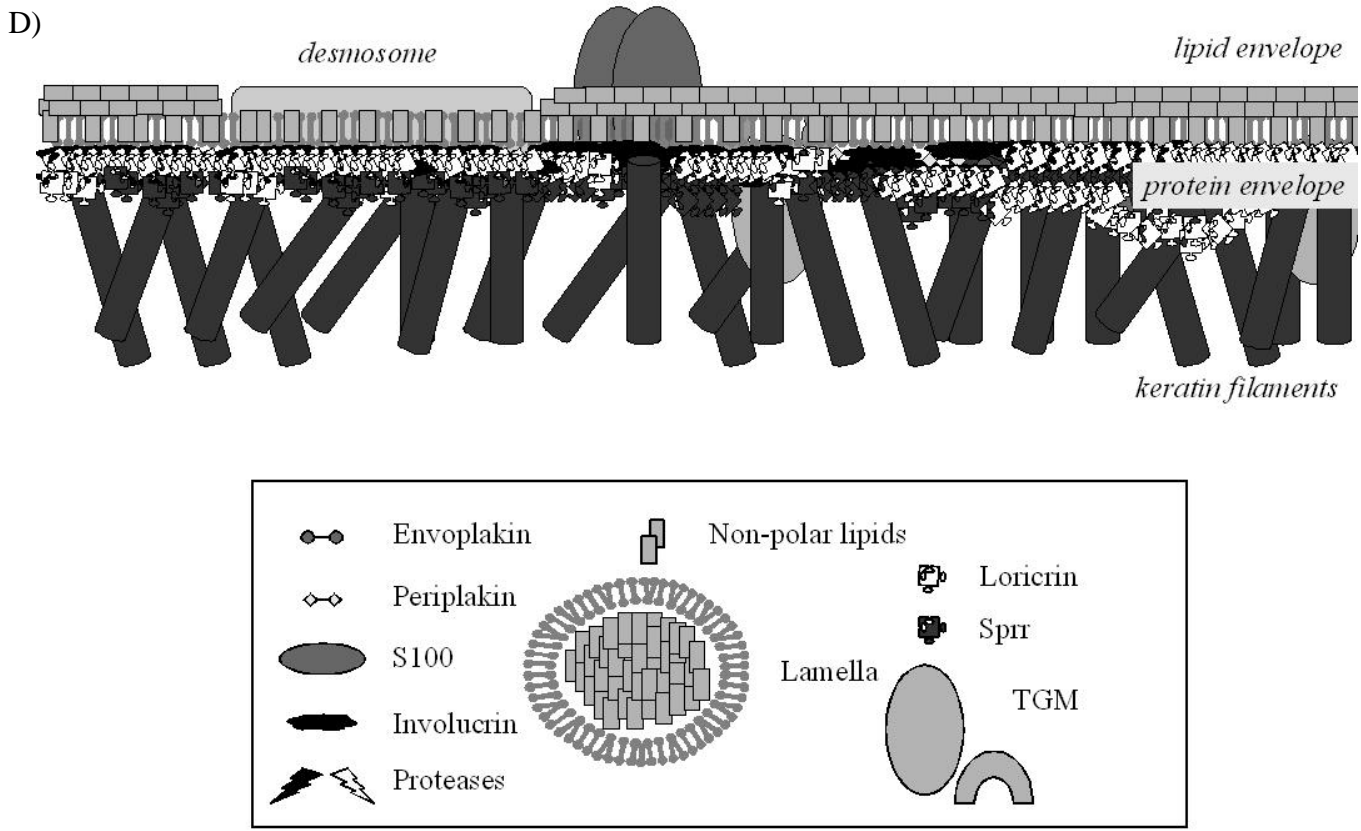

Figure 6    **Appearance and growth of the cornified envelope**. A.- keratinocyte entering cornification; B- initiation of cornification: cells express envoplakin and periplakin; C- early cornification: involucrin binds to the plasma membrane, cell forms lamellar bodies and expresses S100 proteins; D- crosslinking: transglutaminases oligomerize precursors of the cornified envelope and incorporate them into the plasma membrane; E- late cornification: enucleation and degradation of cell organelles, proteolysis and desquamation.

Another essential element of the cornified envelope is TGM (Figure 6D). There are four transglutaminase isoforms expressed in the epidermis: TGM 1, TGM2, TGM3 and TGM5. TGM3 cross-links proteins to short oligomers and TGM1 connects them to the cornified envelope [61]. Specificity of another epidermal TGM isoform, TGM5, seems to be similar to TGM1 [62] while it can't replace it in crosslinking of lipids and doesn't incorporate itself into the plasma membrane [*rev. in* 63]. Contribution of TGM2 to cornification is not yet proven. Besides its participation in protein crosslinking, TGM1 also crosslinks ω-hydroxy-ceramides to involucrin [64]. During the terminal differentiation, rise of the calcium concentration results in proteolytic cleavage and about 1000 fold activation of TGM1 [65; 66]. Recently obtained data suggest that cathepsin D, an aspartate protease, can be involved in proteolytic activation of TGM1 [67]. In cathepsin D-deficient mice, TGM1 activity was sufficiently reduced, likely due to an abnormal processing of this enzyme.

Members of epidermal differentiation complex that have sequence similarities such as glutamine- and lysine-rich tracts become good substrates for TGM (Figure 6D). After $Ca^{2+}$-dependent activation, TGM1 incorporates itself into the membranes and links to the cornified envelope other proteins like envoplakin, periplakin and involucrin [68]. Shortly, members of small proline rich proteins, Sprr, become covalently bound to involucrin and envoplakin. Sprrs are very important for cornification due to their ability to form back links between larger proteins [*rev in* 69 and 70]. Then, the complex of cross-linked proteins spreads across the inner surface of the plasma membrane, penetrates the desmosomes and binds to the intermediate filaments. The lamellar bodies fuse with the plasma membrane and ω-hydroxy ceramides appear in the cytoplasm. Since they become available to TGM1, it links them to the protein layer. At the later stages of keratinocyte differentiation, when the cornified envelope becomes stable and insoluble, it also recruits late-differentiation proteins: loricrin, the major component of the cornified envelope that composes about 70% of the total crosslinked protein and repetin (Figure 6D) [68]. Surprisingly, the epidermis of loricrin knockout mice looked almost normal [71] as it believed due a compensatory effect that appears in upregulation of the certain Sprr proteins and repetin. Interestingly, the knockout mice that lacked an expression of involucrin [72] and envoplakin [73] also did not have serious impairments in the epidermis. On the other hand, targeting structural proteins of the cornified envelope, desmosomal and cytoskeletal proteins often led to severe pathology and disturbance of barrier function. This was demonstrated numerous times by a targeted mutation or an ablation of keratins (K10, K2e and K1), desmoplakin, desmoglein and desmocollin-1 [*rev. in* 63].

## Role of proteolysis in differentiation

Regulation of the protease activity in cells and tissues is a crucial element of their proper functioning that relays on a fragile balance of multiple factors and can be easy destroyed by a sudden shift at wrong place and time. In the most cases, proteolytic enzymes are carefully isolated from their potential targets. Besides, the level of protease inhibitors expressed by the cells is usually sufficient to prevent an occasional leak of endogenous or deactivate exogenous (microbial) proteases. Protease inhibitors may also reside in the intracellular space to keep the secreted enzymes neutralized. In *stratum corneum*, they are represented by antileukoproteinase (ALP) known also as secretory leukocyte protease inhibitor (Slpi); elafin, formerly known as SKALP, skin-derived-antileukoprotease; lymphoepithelial Kazal-type 5 serine protease inhibitor (LEKTI) and two specific cysteine protease inhibitors: cystatin and cystatin M/E. ALP and elafin inhibit the stratum corneum chymotryptic enzyme, SCCE, preventing the detachment of corneocytes from human plantar epidermis [74]. LEKTI (protein product of the gene SPINK-5) targets SCCE and SCTE, stratum corneum tryptic enzyme. Cystatin expression was detected in the spinous layer [75]. During the cornification, cystatin interacts with epidermal TGM and becomes crosslinked to the cornified envelope as one of the minor components [69] Expression of cystatin M/E is restricted to the stratum granulosum [76]. Mutations in this gene impair desquamation and may lead to life-threatening conditions [74, 77, 78].

In this context, cornification can be considered as a serious challenge for the cell antiproteolytic defense. Gradual disruption of cell organelles due to the rapid growth of the cornified envelope releases proteases from their storages. Interfusion of different compartments, particularly lysosomes and cytoplasm, causes a release of potentially disruptive enzymes and involves in the degradation the proteins those function was to suppress it. Indeed, the granular layer of the epidermis is stained positively for a variety of proteases that normally do not appear in the cytoplasm including the lysosomal proteolytic enzymes [79]. On the other hand, the calcium influx decreases the ratio of protease inhibitors to other cytoplasmic proteins. Finally, degradation of the nucleus contributes to the shift. It shouts down the gene expression and it leaves the cell without *de novo* synthesized protein in conditions when it faces a great variety of fully-activated proteolytic enzymes.

Interestingly, two major proteases of *stratum corneum* SCCE/KLK7/hK7 and SCTE/KLK5/hK5 together can destroy three major components of the corneodesmosomes: DSC1, DSG1 and CDSN [80-82]. These enzymes belong to kallikrein family of serine proteases. Their expression starts in suprabasal keratinocytes where their inactive precursors undergo a processing by an unidentified trypsin-like protease [83; 84]. In *stratum corneum*, these enzymes appear in the intercellular spaces [85; 86] suggesting their involvement in the desquamation.

Previously performed analysis of protease inhibitors expressed in the skin, especially the studies carried on the transgenic mice led to the surprising conclusion that other proteolytic enzymes also take a part in cornification and desquamation. This group of enzymes includes cathepsins C and D [87], L and L2 [88], legumain and serine proteinase MT-SP1. Some authors question a sufficiency of cathepsin E for desquamation while even they do not deny its "minor" contribution to the digestion of desmosomes [87].

Cathepsin C (CTSC) is a lysosomal cysteine protease that participates in degradation of the intracellular proteins. Cathepsin C seems to be essential for processing of keratins and contributes to the digestion of desmosomes [89]. Besides, it is important for establishing and maintenance of the structural organization in the tissues that surround the teeth [90]. Mutations in cathepsin C are associated with Haim-Munk and Papillon-Lefevre syndromes [91] that appear in severe early onset periodontitis, premature tooth loss, and thickening and scaling of the skin on palms and soles. Besides, patients with Haim-Munk syndrome (HMS) also have some other clinical features such as nail deformities, acroosteolysis. Surprisingly, CTSC-deficient mice do not resemble the human disorders and do not have severe abnormalities in teeth and skin [92].

Cathepsin D (CSTD) is a lysosomal enzyme and an aspartatic protease. In the epidermis, CSTD performs the final stage of desquamation [87]. It is also important for activation of TGM1 [67]. Ablation of cathepsin D in the mice sufficiently reduced TGM1 activity and also decreased the levels of involucrin and loricrin. Besides, it affected the morphology of *stratum corneum*, which was composed of an increased number of corneocyte layers. Cathepsin D- null phenotype had a similarity to the phenotype observed for the human skin disease, lamellar ichthyosis. Surprisingly, these findings also suggested a link between protease activity and regulation of the gene expression while the molecular mechanism underlying this interesting phenomenon still remains uncovered [93].

Cathepsin L (CTSL) is a lysosomal cysteine protease that participates in processing of trichohyalin [94] and profilaggrin [95]. Cathepsin L-deficient mice develop epidermal hyperplasia, acanthosis, hyperkeratosis and periodic hair loss [96]. *Furless* (fs) and *nackt* (nkt) mice with a mutation in Cathepsin L and lock of Cathepsin L activity develop similar phenotype [95; 96].

Cathepsin L2 formerly known as stratum corneum thiol protease (SCTP) [88; 97] and cathepsin V, has 75% sequence homology with cathepsin L. In mice, cathepsin L2 is not found to the time, while in humans, Cathepsin L2 has even higher activity than its closest homolog, cathepsin L. [88]. Surprisingly, the human cathepsin L2 expressed from K14 promoter can compensate the absence of the murine cathepsin L in cathepsin L deficient mice [98].

Legumain is a cysteine protease that hydrolyzes the proteins at Asp and Asn sites. In lysosomes, legumain processes precursors of other cathepsins and legumain deficiency results in accumulation of unprocessed cathepsins B, H and L. In *stratum corneum*, legumain interacts with cystatin M/E and cystatin M/E deficiency is incompatible with the life. The cystatin M/E knockout mice that survived up to 5-12 day had aberrant cornification in hyperplastic and hyperkeratotic epidermis. Moreover, unrestricted legumain causes premature TGM3 processing and abnormalities in processing of loricrin.

Type II transmembrane serine protease Matriptase (MT-SP1) is involved in numerous biological programs such as hair follicle development and epidermal differentiation [99]. Targeted deletion of this enzyme is lethal for the mice due to dehydration and loss of epidermal barrier function. Besides, it disturbs the formation of lipid matrix, impairs morphogenesis of the cornified envelope and affects desquamation. Some of these consequences can be explained by a defective processing of profilaggrin and loss of mature filaggrin monomer and filaggrin S-100 protein. On the other hand ablation of MT-SP1 seems doesn't affect other epidermal proteins or proteolytic processing of epidermal TGM [*rev. in* 100].

Formation of the mature cornified envelope and intercellular lipid layer accomplishes the transition of cells into flat, keratin-filled corneocytes. Instead of degraded subcellular organelles corneocytes become tightly packed by the keratin fibrils that are oriented roughly parallel to the long dimension of the cell. Between keratin fibrils there is a matrix consisting of the remains of keratohyalin. As soon as corneocytes reach the skin surface their connection to the neighbor cells becomes weaker and finally they take out off it.

## Conclusion

Despite the recent progress in our understanding of epidermal differentiation, the mechanism that coordinates gene expression remains unknown. In the other words, list of the differentiation participants that includes multiple categories such as receptors, kinases, transcription factors and the others still can tell us a little of the mysterious motive that enforces all these diverse parts to work together. That could be caused by several reasons. First, we can't be sure that we know all of them, while studies of epidermal differentiation continue for a long period of time. Indeed, role of many genes remains unclear and it is very unlikely that we are going to know soon what most of them are doing in the cell. Secondary, role of some factors can be different from what do we think of them now. Indeed, several proteins that we previously considered as "only regulatory" (ex. transcription factor Clock) demonstrated the enzyme (histone acetyl transferase) activity. *En contra*, other proteins that were initially described as enzymes also appeared as regulatory factors (ex. epidermal protease Cathepsin D that is also a receptor ligand). Third, some parts of the mechanism still can be missing or we previously didn't pay them enough attention. The mechanisms of epigenetic control such as histone modification can be a good example of it. Fourth, we honestly do not know how the previous step of differentiation predetermines the next one. Finally, we should know sufficiently more of the external stimuli that come into the epidermis from outside and push the trigger of the differentiation program. On the other hand, the existence of a common coordinating center for gene expression mentioned above can be proved *via* establishing a new category of the systemic transcription factors. These factors suppose to connect epidermis with the surrounding tissues and rapidly react to the incoming stimuli. Besides, their principle role in the cell would be in preparing the chromatin to the next differentiation step *via* histone modification. The future search for these class of regulatory elements already begun [101] and it expects to be a challenging task.